\documentclass[prl,amsmath,superscriptaddress,twocolumn,showpacs]{revtex4}
\usepackage{bm}
\usepackage{amssymb}\usepackage{amssymb,amsmath}   % need for subequations
\usepackage[dvips]{graphicx}   % need for figures
\usepackage{verbatim}   % useful for program listings
\usepackage{color}      % use if color is used in text
\usepackage{subfigure}  % use for side-by-side figures
\usepackage{hyperref}   % use for hypertext links, including those to external documents and URLs
\usepackage{gensymb}
\usepackage{epstopdf}
\usepackage{natbib}

\usepackage{color}
\definecolor{orange}{rgb}{1,0.5,0}
\definecolor{darkgreen}{rgb}{0,0.5,0}
\definecolor{darkblue}{rgb}{0,0,0.5}
\definecolor{purple}{rgb}{0.35,0,0.35}


\begin{document}
\title{Reentrance and Magnetic-Field-Induced Superconductivity with Kondo Impurities: Bulk and Proximity Systems}

\author{Yuval B. Simons}
\affiliation{Department of Condensed Matter Physics, Weizmann Institute of Science, Rehovot 76100, Israel}
\author{Ora Entin-Wohlman}
\altaffiliation{Also at Tel Aviv University.}
\affiliation{Department of Physics, Ben-Gurion University, Beer
Sheva 84105, Israel} \affiliation{Ilse Katz Center for Meso- and
Nano-Scale Science and Technology, Ben-Gurion University, Beer
Sheva 84105, Israel}
\author{Yuval Oreg}
\affiliation{Department of Condensed Matter Physics, Weizmann Institute of Science, Rehovot 76100, Israel}
\author{ Yoseph Imry}
\affiliation{Department of Condensed Matter Physics, Weizmann Institute of Science, Rehovot 76100, Israel}
\date{\today}
\begin{abstract}
Reentrant behavior is known to exist and magnetic-field-induced superconductivity has been predicted in superconductors with Kondo impurities.
%Magnetic-field-induced superconductivity has been predicted and reentrant behavior has been predicted and observed for superconductors with Kondo impurities.
 We present a simple framework for understanding these phenomena and generalize it to explain the long-standing puzzle of paramagnetic reentrance in %cylindrical
thick proximity systems as due to small amounts of Kondo impurities.

%generalize it, in the appropriate limit, to other similar systems. We use this framework to explain the long standing puzzle of paramagnetic reentrance in cylindrical proximity systems as due to small amounts of Kondo impurities.
\end{abstract}
\maketitle
%
%\pagenumbering{arabic}
%\section{Introduction}
{\bf Introduction.} Magnetic impurities act as pair breakers in a superconductor since their interaction with the electrons breaks the time-reversal symmetry of the two members of each Cooper pair. Many works have dealt with the effect of magnetic impurities on superconductivity and the landmark for these  is Abrikosov and Grokov's calculation of the depression of the critical temperature of a superconductor in the presence of magnetic impurities \cite{AG}. Maki \cite{Maki} expanded this work to include any pair-breaking mechanism in a universal manner through the pair-breaking energy $\alpha$. When the pair-breaking energy is of the order of the critical temperature superconductivity will be completely suppressed.

In 1970 M\"uller-Hartmann and Zittarz \cite{zittartz} expanded Abrikosov and Gorkov's work to include the Kondo effect. They predicted that due to the competition between superconductivity and spin-flip scattering off the Kondo impurities there will be, for certain impurity concentrations, not one but three critical temperatures, that is as the temperature is lowered the superconductor will go into a superconducting state at $T_{C1}$, out of it at a lower temperature $T_{C2}$ and back into it at an even lower temperature $T_{C3}$. This happens because the Kondo impurities' spin-flip depairing is maximal around $T_K$, the Kondo temperature, and falls off far from it \cite{universal}. This reentrant behavior was confirmed experimentally as far as the existence of $T_{C2}$, a reentrance into a normal phase, is concerned \cite{EarlyMota} but the existence of a third transition and the conditions under which it will be observable are still a matter of debate \cite{thirdtransdebate}.

In 1989, in an often overlooked set of articles \cite{scoopers}, Podmarkov and Sandalov predicted the existence of magnetic-field-induced superconductivity in such systems. They predicted that while for small temperatures and magnetic fields superconductivity will be suppressed by spin-flip scattering off the Kondo impurities the application of magnetic field can polarize the impurities thus reducing the spin-flip scattering and restoring superconductivity.

In this work we shall use a simple interpolation in order to expand M\"uller-Hartmann and Zittartz's work to include the effect of a finite magnetic field and thus account for Podmarkov and Sandalov's predictions.  We shall also see why such behavior should appear in other superconducting systems and since the reentrant phenomenon in thick cylindrical proximity systems, which has intrigued theoreticians since its discovery in 1990 by Visani et al. \cite{mota1}, shows such behavior we will suggest that it arises due to a small concentration of Kondo impurities in the measured samples and estimate the amount of impurities needed to explain it.
 
{\bf The Influence of Magnetic Fields on Superconductors with Kondo Impurities.} 
%For a bulk superconductor the effect of pair breakers is universal, i.e. does not depend on the nature of the specific pair breaking mechanism \cite{Maki}, and was first calculated by Abrikosov and Gorkov \cite{AG}.
The Abrikosov-Gorkov formula for the depression of the critical temperature due to magnetic impurities is
\begin{align}
\label{AGeq}
\ln \left( \frac{T_C}{T_{C0}}  \right)=\Psi\left( \frac{1}{2} \right)-\Psi \left( \frac{1}{2} +\frac{\alpha}{2 \pi T_C}\right)\ ,
\end{align}
with $T_{C0}$ being the transition temperature without pair breaking, $T_C$ the transition temperature, $\alpha$ the pair-breaking energy,  and $\Psi$ the digamma function. The critical temperature is completely suppressed when %the pair breaking energy, 
$\alpha$ reaches a critical value $\alpha_{cr} \equiv \pi/2 \exp(-\gamma) T_{C0}=0.882 \, T_{C0}$ with $\gamma$ being the Euler constant.%As we can see in Fig. \ref{AG}, 

M\"uller-Hartmann and Zittartz  \cite{zittartz}, building on the works of Nagaoka \cite{Nagaoka} and Suhl \cite{Suhl}, arrived at an approximate form for the temperature dependence of the pair-breaking energy for small concentrations of Kondo impurities
\begin{align}
\label{NS}
\alpha(T)=\alpha_{\text{max}} \frac{\pi^2 S (S+1)}{\pi^2 S (S+1) + \ln^2(T/T_K)}\ ,
\end{align}
with $T_K$ being the Kondo temperature, $S$ being the spin of the magnetic impurities and $\alpha_{\text{max}}=n_s/ 2 \pi \nu$ with $n_s$ being the density of the impurities and $\nu$ being the density of states at the Fermi level. 
%As can be seen from this formula, the pair breaking is maximal around the Kondo temperature when spin-flip scattering of electrons from the impurities is most likely. 
At low temperatures, $T \ll T_K$, this formula, known as the Nagaoka-Suhl formula, ceases to be valid and more refined methods have to be used (e.g. \cite{universal}).

\begin{figure}[ht]
\begin{center}
\includegraphics[width=3in]{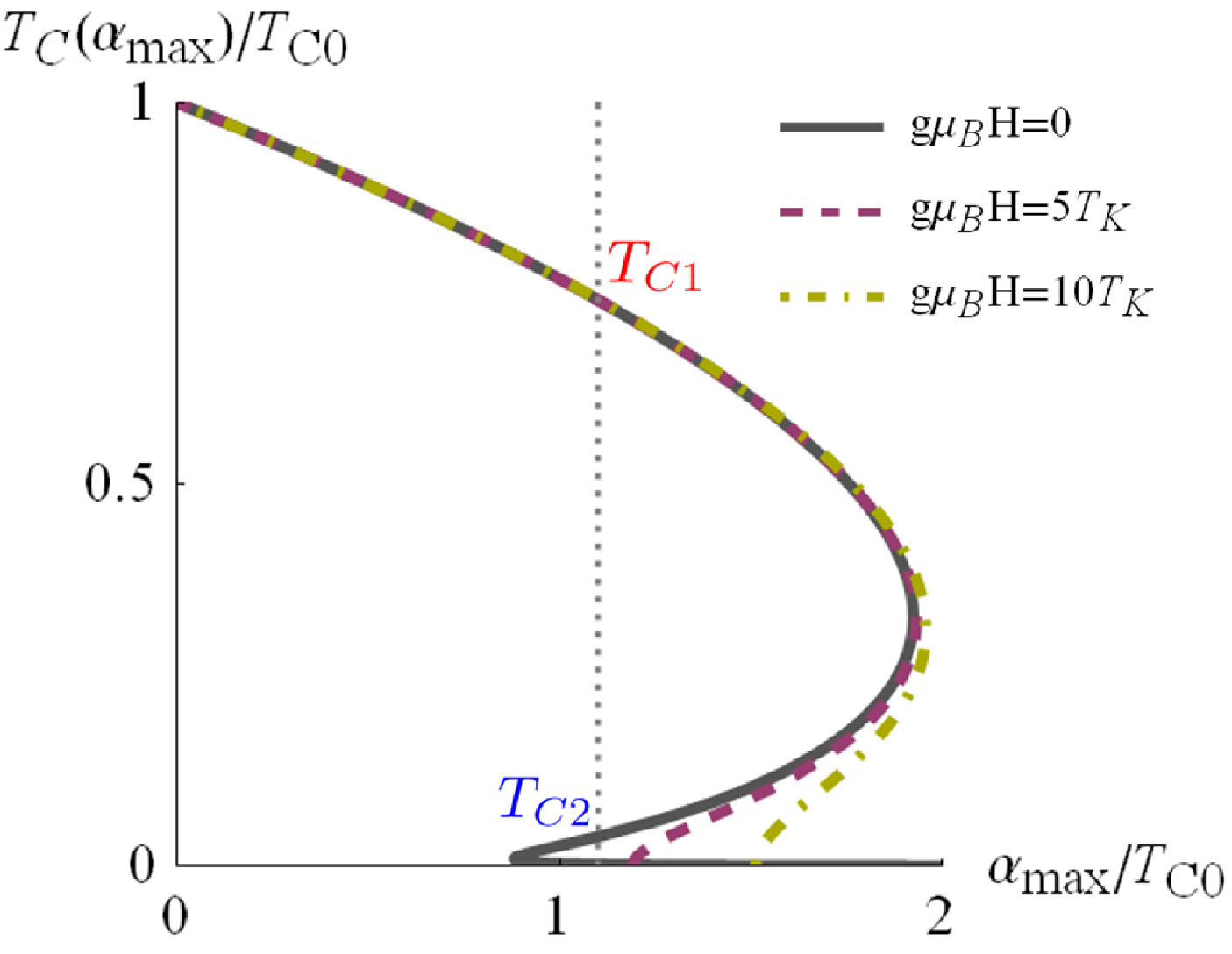}
\caption{The critical temperature of a bulk superconductor in the presence of Kondo impurities, $T_C(\alpha_{\text{max}})$, versus the maximal pair-breaking energy of the Kondo impurities $\alpha_{\text{max}}$. Drawn from the self-consistent solution of Eqs. \eqref{AGeq} and \eqref{NSTH} for $T_K=T_{C0}/100$, $S=1/2$ and three values of the Zeeman energy $g \mu_B H = 0$, $5 T_K$ and $10 T_K$ (in gray, dashed purple and dot-dashed yellow). The reentrant range for which there exists two or three transition temperatures is clearly visible and the magnetic field suppresses this reentrant behavior. The dotted gray line is the value of $\alpha_{\text{max}}$ for which the dependence of the transition temperatures $T_{C1}$ and $T_{C2}$ (shown in red and blue measured in units of $T_{C0}$) on the magnetic field is plotted in Fig. \ref{phase} . The third transition temperature, $T_{C3}$, which might exist at temperatures well below $T_K$ is beyond the scope of this work.
\label{triple}}
\end{center}
\end{figure}

M\"uller-Hartmann and Zittartz obtained the impurity dependence of the superconductor's critical temperature by plugging $\alpha(T_C)$ from Eq. \eqref{NS} into Eq. \eqref{AGeq} and solving for $T_C$ thus resolving self-consistently the interplay between superconductivity and Kondo impurities' pair breaking. The resulting dependence shows a striking reentrant behavior for certain impurity concentrations, see Fig. \ref{triple}, which was confirmed experimentally \cite{EarlyMota}. %but not the existence of the third transition. %, or more accurately the low temperature limit of $T_C(n_s)$, remains a matter of debate \cite{thirdtransdebate}.\\
%Kondo impurities break Cooper pairs by spin-flip scattering, that is events in which an electron tunnels into the impurity with spin up (down) and tunnels out with spin down (up) so the pair breaking energy is determined by the spin-flip rate $\alpha=1/\tau_{sf}$ with $\tau_{sf}$ being the spin flip rate. 

%The Zeeman interaction polarizes the Kondo impurities and so affects their pair-breaking ability. When the Zeeman energy due to a finite magnetic field is much larger than the temperature we can replace the temperature in Eq. \eqref{NS} with the Zeeman energy $g \mu_B H$  \cite{mag}, with $g$ being the Zeeman g-factor, $\mu_B$ the Bohr magneton and $H$ the magnetic field.

As long as the Zeeman interaction is the dominant effect of the magnetic field on the system we can account for the effect of a finite magnetic field by using the interpolation \cite{interpol}
\begin{align}
\label{NSTH}
\alpha(T, H)=\alpha_{\text{max}} \frac{\pi^2 S (S+1)}{\pi^2 S (S+1) + \ln^2(\sqrt{T^2+(g \mu_B H)^2}/T_K)}
\end{align}
with $g$ being the Zeeman g-factor, $\mu_B$ the Bohr magneton, and $H$ the magnetic field. This equation interpolates between Eq. \eqref{NS}, which is the $g \mu_B H \ll T$ limit, and the $g \mu_B H \gg T$ limit where the Zeeman energy, $g \mu_B H$, replaces the temperature in that equation \cite{mag}. The magnetic field's dominant effect on the system is the Zeeman interaction when it is much smaller than the critical field of the superconductor and other pair-breaking effects \cite{Maki} are small compared to $\alpha_{\text{max}}$.

There have been several works regarding the crossover between these limits \cite{crossover} but this simple interpolation, which disregards the pair-breaking energy's dependence on the electrons' energy, allows for an easy analysis of the system's behavior while giving the same qualitative results as other, more sophisticated methods \cite{scoopers} and allowing for easy generalization to other, similar systems (see below).

%The Zeeman interaction polarizes the Kondo impurities and so affects their pair-breaking ability and when the Zeeman energy due to a finite magnetic field is much larger than the temperature then the pair-breaking energy can be approximated by \cite{mag}
%\begin{align}
%\label{NSTH}
%\alpha(T, H)=\alpha_{\text{max}} \frac{\pi^2 S (S+1)}{\pi^2 S (S+1) + \ln^2(g \mu_B H/T_K)}\ ,
%\end{align}
%with $g$ being the Zeeman g-factor, $\mu_B$ the Bohr magneton and $H$ the magnetic field.

%For the intermidiate regime  between the $g \mu_B H \ll T$ and the $g \mu_B H \gg T$ limits we can use the interpolation \cite{interpol}
%\begin{align}
%\label{NSTH}
%\alpha(T, H)=\alpha_{\text{max}} \frac{\pi^2 S (S+1)}{\pi^2 S (S+1) + \ln^2(\sqrt{T^2+(g \mu_B H)^2}/T_K)}.
%\end{align}
%There have been several works regarding the crossover between these limits \cite{crossover} but this simple interpolation, which disregards the pair-breaking energy's dependence on the electrons' energy, allows for an easy analysis of the system's behavior while giving the same qualitative results as other, more sophisticated methods \cite{scoopers} and allowing for easy generalization to other, similar systems (see below). Notice that Eq. \eqref{NSTH} can only be used as long as the Zeeman interaction is the dominant effect of the magnetic field on the system. Specifically, $H$ must be smaller than the critical field of the superconductor,  and other pair-breaking effects \cite{Maki} must be small compared to $\alpha_{\text{max}}$.

\begin{figure}[ht]
\begin{center}
\includegraphics[width=3in,clip=]{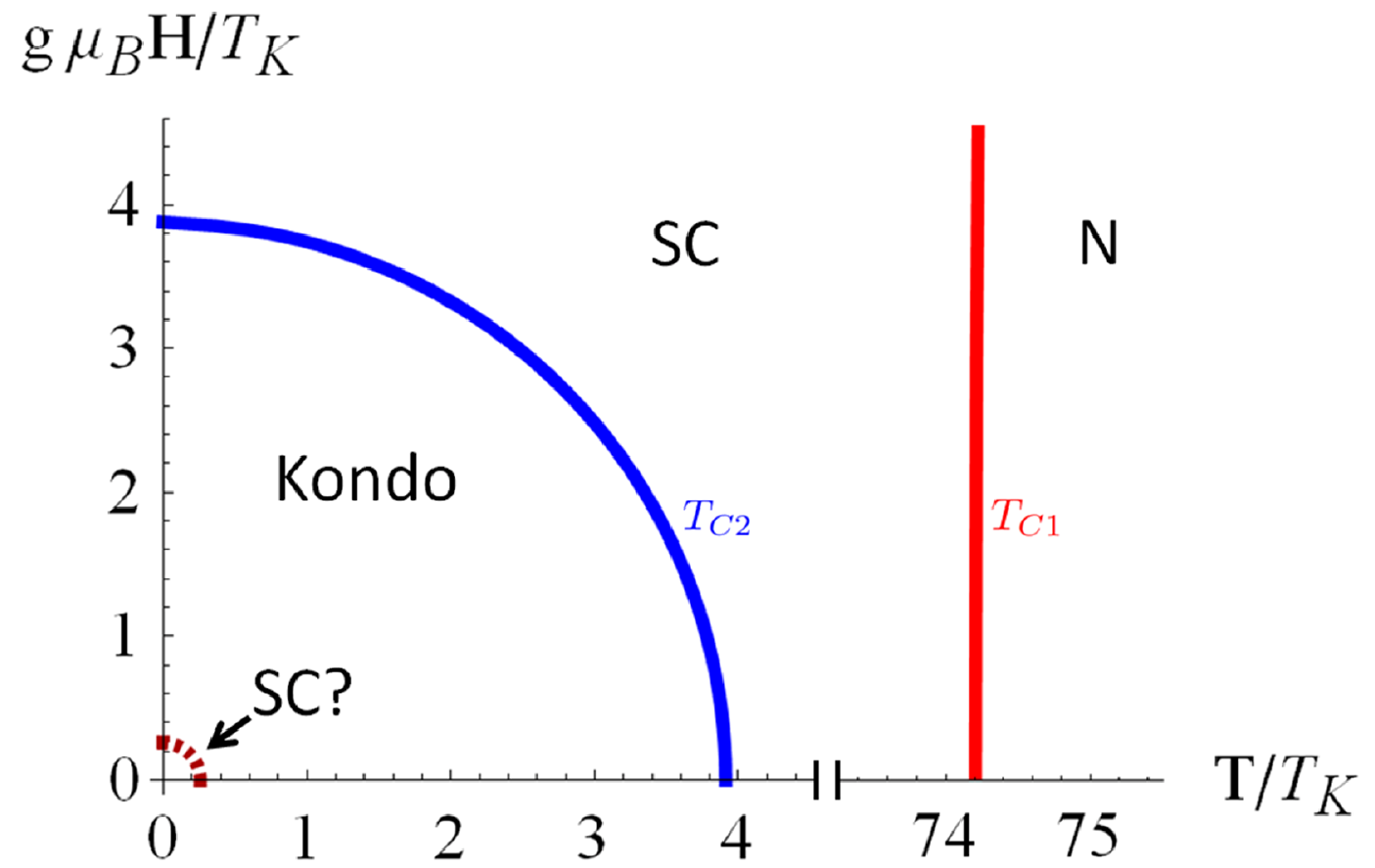}
\caption{Phase diagram of a bulk superconductor with Kondo impurities in a small magnetic field,  drawn from the self-consistent solution of Eqs. \eqref{AGeq} and \eqref{NSTH}     %Shown here   
for $T_K=T_{C0}/100$ and $\alpha_{\text{max}}=1.1 T_{C0}$. At $T=T_{C1}\simeq0.74 T_{C0}$ the system has a phase transition from a normal phase into a superconducting one which at $H \ll H_{C2}$ is almost field independent. At a low temperature, $T_{C2}\simeq 4 T_K= 0.04 T_{C0}$, the system undergoes a second phase transition into a non-superconducting phase dominated by spin-flip scattering from the Kondo impurities which may be suppressed by a magnetic field of the order of $g \mu_B H = T_{C2}$. Thus the application of magnetic field in this phase can induce superconductivity. At even lower temperatures and magnetic fields ($T, g \mu_B H \ll T_K$) there might exist a transition back into a superconducting state.
\label{phase}}
\end{center}
\end{figure}

We can now insert Eq. \eqref{NSTH} into Eq. \eqref{AGeq}, solve for $T_C$ and obtain the dependence of the critical temperature on both impurity concentration and magnetic field as seen in Fig. \ref{triple} and the phase diagram of our system as can be seen in Fig. \ref{phase}. For small temperatures and magnetic fields, $T,g \mu_B H \simeq T_K$, the system stops superconducting and goes into a phase dominated by spin-flip scattering %from 
off the Kondo impurities. For even smaller temperatures and magnetic fields there might exist another region of superconductivity but since the Nagaoka-Suhl approximation is insufficient in that region we cannot contribute to the debate over its existence. However, we can mention that this extra region of superconductivity's extreme sensitivity to magnetic field might be a contributing factor in its experimental elusiveness.

In the Kondo phase, application of a magnetic field of the order of $T_K/ \mu_B$ will cause the appearance of superconductivity by suppressing the Kondo effect. In this magnetic-field-induced superconductivity the Meissner effect will act to screen the magnetic field from the impurities but since magnetic field is necessary for superconductivity a balance must be obtained and we shall therefore expect to see reduced Meissner screening which will increase at higher fields. There have already been several reports of systems with magnetic-field-induced superconductivity \cite{mfiSC} including superconducting systems with magnetic impurities but this type of behavior with both reentrance and magnetic-field-induced superconductivity is yet to have been measured.   %\\ Superconductors with Kondo impurities can show both reentrant behavior in temperature and magnetic field induced superconductivity.\\
%We can gain good insight into this system by looking at a graphical solution of the equations for $T_C$, Eqs. \eqref{AGeq} and \eqref{NS}, in Fig. \ref{graphsol}. The transition temperatures are the intersections of the two graphs $T_C(\alpha)$ from Eq. \eqref{AGeq} and $\alpha(T)$ from Eq. \eqref{NS}.\\ 
%\begin{figure}[ht]
%\begin{center}
%\includegraphics[width=3in]{graphicsloution.pdf}
%\caption{The shape of $T_{C}(\alpha)$ (in red), the critical temperature as a function of the pair breaking energy from Eq. \eqref{AGeq}, and $\alpha(T)$ (in dashed blue), the pair breaking energy as a function of temperature from Eq. \eqref{NS}. Also seen is $\alpha_{cr}$ (in dotted mustard), the critical pair breaking energy. Superconductivity will be seen at the range in which $\alpha(T)$ is below $T_{CN}(\alpha)$ and so the the crossings of $\alpha(T)$ and $T_{CN}(\alpha)$ represent critical temperatures. For a small enough $T_K$ there can be values of $\alpha_{\text{max}}=n_s/ 2 \pi \nu$, the peak of the pair breaking energy, for which reentrance will occur i.e. there will be multiple critical temperatures.
%\label{graphsol}}
%\end{center}
%\end{figure}

In the limit $T_K<T_C \ll T_{C0}$ we can obtain approximate analytic expressions for the reentrant temperature and the range of impurity concentrations leading to reentrance which will be valid not only for bulk superconductors but also for other superconducting systems.  At $T_C \ll T_{C0}$ Eq. \eqref{AGeq} reduces to
\begin{align} 
\label{alpha}
\alpha=\alpha_{cr}-O(T_C^2/T_{C0})\ ,
\end{align}
and since $T_K \ll T_{C0}$ we can use the temperature-dependent $\alpha$ from Eq. \eqref{NSTH} and set $\alpha(T_C)=\alpha_{cr}$. At zero field this gives us that as long as the maximal spin-flip induced pair breaking exceeds the critical pair breaking energy, that is
\begin{align} \label{cond1}
\alpha_{\text{max}}>\alpha_{cr} ,
\end{align}
there exists a transition from a superconducting phase into a non-superconducting phase %dominated by spin-flip scattering from the Kondo impurities
 when the pair breaking exceeds $\alpha_{cr}$ at the temperature
\begin{align} 
\label{TC2}
T_{C2}(H=0)=T_K \exp\left(\pi \sqrt{S \left(S+1\right)\left (
\alpha_{\text{max}}/\alpha_{cr}
%\frac{\alpha_{\text{max}}}{\alpha_{cr}}
-1\right)}\right).
\end{align}
When $T_{C2}\sim T_{C0}$ this approximation  breaks down but that is exactly the limit where we expect reentrance to cease to exist so we can use the condition $T_{C2}<T_{C0}$ with the above expression to obtain an upper bound for the reentrant range
\begin{align} \label{cond2}
\alpha_{\text{max}}<\alpha_{cr} \frac{\pi^2 S (S+1) + \ln^2(T_{C0}/T_K)}{\pi^2 S (S+1)}\ .
\end{align}
%and so from 
Using Eqs. \eqref{cond1} and \eqref{cond2} we obtain %the expression
\begin{align}
2 \pi \nu\alpha_{cr}<n_s<2 \pi \nu \alpha_{cr}\frac{\pi^2 S (S+1) + \ln^2(T_{C0}/T_K)}{\pi^2 S (S+1)} \label{range}
\end{align}
for the range of impurity concentrations leading to reentrance.

Repeating this %exercise 
procedure at a finite magnetic field we find that   %immediately see that 
\begin{align} \label{TC2H}
T^2_{C2}(H)+(g \mu_B H)^2=T^2_{C2}(H=0)
\end{align}
%and we see that this is where the shape of the interpolation, Eq. \eqref{NSTH}, comes into effect and it is very similar to the shape of $T_{C2}(H)$ as determined by a more detailed calculation \cite{scoopers}. 
and we see that as the field increases $T_{C2}(H)$ decreases. When $T_{C2}(H)$ goes below the sample's temperature the system will go out of the phase dominated by spin-flip scattering from the Kondo impurities and become superconducting, see Fig. \ref{phase}.

The shape of Abrikosov and Gorkov's formula for $T_C(\alpha)$, Eq. \eqref{AGeq}, does not come into play in Eqs. \eqref{TC2}, \eqref{range} and \eqref{TC2H}, only the value of $\alpha_{cr}$ does. Since the shape of spin-flip induced Kondo impurity pair breaking is universal \cite{universal} then these equations should hold for any system where a critical pair-breaking energy destroys superconductivity and we should expect to see the behavior described above qualitatively in such systems even when Eq. \eqref{AGeq} is not valid. Examples of such systems 
are thin superconducting films, dirty superconductors and proximity induced superconductors (Kondo impurity induced reentrance has already been predicted for thin proximity systems \cite{kaiser}).

{\bf Paramagnetic Reentrance in Proximity Systems.} 
We now turn our attention to a system that shows a very similar behavior to the one we have just described, a behavior which has baffled theoreticians for over twenty years. The system in question is the %cylindrical 
thick proximity cylinders measured by Visani et al. and subsequent experiments \cite{mota1, mota94, mota96, mota2000} which show both reentrant %behavior
 and magnetic-field-induced superconducting behavior, the two hallmarks of the 
%work 
picture described above.

In 1990, Visani et al. \cite{mota1} measured the magnetic response of relatively clean thick proximity cylinders of superconducting material (Nb, Ta) coated with normal metal (Ag, Cu). As the samples were cooled the superconductor started to show full diamagnetic screening, that is the Meissner effect, at $T \simeq T_{CS} \sim 10 K$, the critical temperature of the bulk superconductor, as expected. Around $T \simeq T_A \sim 1 K$, the Andreev temperature (see definition below), the samples showed, as expected, the proximity effect, that is the normal metal started to show the Meissner effect as well.

In some samples at a lower temperature ($\sim$20 mK) a paramagnetic response came into play that tended to cancel the diamagnetic screening. This response increased gradually as the temperature was decreased until it saturated at $T \sim$  1 mK\cite{mota2000}. In some of the samples the normal metal's diamagnetic response was entirely canceled out by the paramagnetic effect and in one sample the paramagnetic effect was reported to cancel out also the superconductor's signal as well. A small magnetic field, of the order of 20 Oe, was shown to be enough to destroy the paramagnetic signal and return the sample to a screening fraction close to that which would have been expected if it wasn't for the reentrant effect. These measurements, which have stimulated a very lively discussion among theoreticians \cite{BI,  BBN,  FBB, proFBB, antiFBB1, antiFBB2, MakiHaas}, can be simply understood, in light of the theory presented above, as due to the existence of Kondo impurities in the normal metal.

For a thick proximity sample the relevant transition temperature is the Andreev temperature \cite{motabelzig}, $T_A= v_F / 2 \pi d \ll T_{CS}$ with $d$ being the thickness of the normal metal and $v_F$ being the Fermi velocity. It is around this temperature that the normal side of the sample will start showing the Meissner effect and it plays the role of $T_{C0}$ when applying our results to this system and so $\alpha_{cr} \sim T_A$. The field in which the supposed Kondo region breaks down is 20 Oe which correlates to a Zeeman energy of $\sim 5 \text{ mK}$,  %which is 
of the same order of the reentrant transition temperature as expected from Eq. \eqref{TC2H}. We can therefore estimate the Kondo temperature of the impurities in question to be $\sim$1 mK.

With these parameters we can now use Eq. \eqref{range} to assess the range of impurity concentrations needed to explain this phenomenon as
\begin{align}
70 \text{ ppm} < n_s < 430 \text{ ppm}\ ,
\end{align}
where we set $S=1/2$, $T_{C0}=T_A=500$ mK, $T_K =1$ mK, $\nu_{Ag}\simeq 0.26 \text{ atom}^{-1} \text{eV}^{-1}$ and $\alpha_{cr}=T_A$ in order to appreciate the reentrant range. The sample in which the Meissner effect disappeared from the superconducting metal itself must have had a higher amount of impurities, around $2 \pi \nu T_{CS} \sim 1000\text{ ppm}$, such that it managed to affect not only its host metal but the adjacent superconductor as well. These amounts of impurities are very small and might arise from various sources during the samples' preparation and measurement.

Past research into this reentrant phenomenon has led to two main approaches.  One approach by Fauch\'ere et al. \cite{FBB} attributed this phenomenon to repulsive net electron-electron interactions in the normal metal. While attracting much theoretical attention \cite{proFBB,antiFBB1,antiFBB2,MakiHaas}
recent experimental results \cite{FBBdeath, PrivateCommBelzig} indicate that such repulsive interactions do not exist in the relevant metals. Another approach by Bruder and Imry \cite{BI} suggested the phenomenon might be due to the contribution of glancing states. Bruder and Imry took the magnetic field to be constant in space in their argument but calculations that treated the magnetic field self-consistently with the Meissner current have shown that the effect of these states is too small to explain this phenomenon \cite{BBN, me}.

The approach presented here explains, at least qualitatively, this phenomenon and can easily be verified experimentally either by characterizing the impurities in the original samples used or by measuring the magnetic response of thick proximity samples with controlled amounts of Kondo impurities. This approach does not rely on geometry and should apply equally to proximity cylinders and slabs.
 %There is also obviously a need to carefully work out the mathematical details involved in a thick, clean proximity system with Kondo impurities in order for a detailed comparison with experiment to be possible.

In summary, we have presented a simple approach for understanding reentrant and magnetic-field-induced superconducting behavior in systems with Kondo impurities and have used it to show that the existence of such impurities can explain the paramagnetic reentrant effect which has been measured in proximty samples. 
%Using a suitable interpolation between two known limits we have managed to characterize the behavior of a superconductor with Kondo impurities under finite but small magnetic fields for small Kondo temperatures $T_K\ll T_{C0}$. At small temperature and magnetic field the Kondo effect dominates over the superconductivity thus resulting in reentrant behavior and magnetic-field induced superconductivity. At even smaller temperature and field a transition back into superconductive behavior may appear.\\
%We have drawn a parallel between the behavior of superconductors with Kondo impurities and the reentrant phenomenon which appears in cylindrical proximity samples at low temperatures. It seems that the existence of a small amount of Kondo impurities, around 100ppm, with a Kondo temperature of ~1mK in the normal side of these samples explains this phenomenon quite well.
\begin{acknowledgments}
We thank W. Belzig and C. Bruder for helpful discussions. OEW acknowledges the support of the Albert Einstein Minerva Center for Theoretical Physics, Weizmann Institute of Science. This work was supported by the BMBF within the DIP program, BSF, ISF and its Converging Technologies Program.
%We acknowledge support
%from WHO EXACTLY PAYS MY SALARY?.
\end{acknowledgments}

\end{document}